\documentclass[a4paper]{article}
\usepackage{amssymb}
\usepackage{bm}
\usepackage{graphicx}
\usepackage{amsmath}
\begin{document}
%\loadspellchecklist[en]
%\setupspellchecking[state=start]
%\mainlanguage[en]
\title{The Particle-Field Theory and Its Relativistic Generalization }
\author{{ Fatemeh Ahmadi$^1$\thanks{e-mail: f.ahmadi@bzte.ac.ir; fatemehs.ahmadis@gmail.com;
} ,  \,\,\,Afshin Shafiee$^{2,3}$\thanks{ Corresponding
author, e-mail:
shafiee@sharif.edu}
%\thanks {This research was in part supported by a grant from IPM (No. 88460118)}
}\\[0.4cm]
{ \em $^{1}$ \small Department of Engineering Sciences and Physics  , }\\
{\em  \small Buein Zahra Technical University, Ghazvin, Iran.}\\
{ \em $^{2}$ \small Research Group on Foundations of Quantum Theory and Information,} \\
{\em \small Department of Chemistry, Sharif University of Technology,}\\
{\em \small  P.O. Box 11365-9516, Tehran, Iran}\\
{\em  $^{3}$  \small School of Physics, Institute for Research in Fundamental Sciences}\\
{\em \small  (IPM), P.O. Box 19395-5531,
Tehran, Iran}}
\date{}
\maketitle
\begin{abstract}
As a serious attempt for constructing a new foundation for describing micro-entities from a causal standpoint,
it was explained before in \cite{shafiee, shafiee2, shafiee3} that by unifying the concepts of information, matter and energy, each micro-entity is assumed
to be composed of a probability field joined to a particle called a particle-field or PF system.\\
% In the [2], some odd quantum phenomena (including the tunneling effect, the measurement problem and EPR theorem)
%were reconsidered from the standpoint of the new theory.\\
\indent
In this essay,  the relativistic generalization of the PF theory has been considered. The equation of motion for the PF system is derived in a form which is Lorentz-invariant. Moreover, based on constitutional
similarities to classical equations of motion, a well-defined  relativistic time-independent Schr\"{o}dinger equation
is derived, which is one of our main achievements in developing a micro-relativistic physics of PF theory.
%We derive relativistic Schr$\ddot{o}$dinger equation for the case of mass-dependent and mass-independent potentials.
This relativistic Schr\"{o}dinger equation is solved  for a relativistic micro-particle in one-dimensional box to find  its eigenstate and energy spectrum.
\end{abstract}
%\pacs{PACS numbers:03.65.Ca; 03.65.w;03}
%\maketitle
\section{Introduction}
"Quantum mechanics is certainly imposing. But an inner voice tells me that it is not yet the real thing. The theory says a lot, but does not really bring us closer to the secret of the ‘Old One.’ I, at any rate, am convinced that He is not playing at dice"\cite{einstein}.\\
\indent  This view on quantum mechanics is now shared by a large number of scientists spanning the entire spectrum of physics, from pure theoretical ones to
cutting edge experimenters.
%Quantum mechanics is a mathematical tool which enables us to make some predictions about the probability of obtaining different outcomes in repeated experiments. This theory provides with us nothing about the individual results, even at the observational level of perception.
There is no inclusive consensus among the physicists and the philosophers of science about the meaning of quantum theory and the way one is preferred to look at quantum world.
There are many weird quantum phenomena that the most important of them are the measurement problem, the EPR paradox \cite{epr} and the quantum interference phenomenon described by the famous double-slit experiments.
No generally accepted variant of quantum theory has been provided up to now to explain these puzzling phenomena. \\
\indent
To be involved in such an important business,  in recent years a new foundation for describing micro-events from a deterministic causal standpoint is formulated, in which a micro-entity is supposed  to be an allied particle-field system, instead of composing of a particle and (or) a field (wave) \cite{shafiee,shafiee2,shafiee3}.
It has been explained in the first essay of this series that in the microworld, one encounters an unified concept of information, matter and energy \cite{shafiee}.
In this new approach,the principles of realism and causality based on the classic-like equations of motion are presumed and the meaning of  wave function is explained  to explain why its form (according to Born postulate) determines the probability density of finding a particle somewhere in space. One can also see a clear depiction for some weird quantum phenomena such as the tunneling effect, double slit experiment and the so-called   EPR thought experiment \cite{shafiee2,shafiee3}.\\
\indent
From a more fundamental point of view, this theory provides us with a new formulation of quantum phenomena based on a unified concept of information (by which we gain knowledge about the possible locations with in which a particle can be found), matter (characterized by the existence of a particle) and energy (attributed to the whole particle-field). This is somehow similar to the special relativity theory in which the concepts of matter and energy are joined  to one concept in the relativistic domain.\\
\indent
So, we suppose that there is a field associated with a particle which together form a unit entity called "particle-field" (PF). The field has a mathematical representation which determines the spatial distribution of the entire system, i.e., it determines the probability of finding  the particle within a definite interval of space, when one measures its position. Moreover, both the particle and the field satisfy deterministic equations of motion, but the field has no independent identity without the existence of the particle. We are not able to see an independent particle directly without any intervention. A PF system is indeed an extended notion of  particle which its nature differs from a classical particle because the particle shares some of its energy with its surrounding space \cite{shafiee}.\\
\indent
%In the second paper of this series, some weired quantum phenomena such as tunneling effect, measurement problem, EPR paradox and double slit experiment has been explained from the standpoint of this new theory. \\
%
\indent
 The other side of non-classical behavior of a system is that ,  the world is not only quantized but also it  is relativistic. It is a four dimensional universe in which the laws of physics obey the principles of relativity. So, we need to show that this new theory fulfills the requirements of the theory of relativity. For this requirement we introduce a unified concept of spacetime which shows the infinitesimal distance  of two separated points based on the defined entities in the PF theory and show that this element  is invariant under Lorentz transformation.\\
\indent
 Moreover,  we obtain the relativistic time-independent Schr\"{o}dinger equation which is one of the achievements of the relativistic generalization of this theory. Here, we consider a non-relativistic potential and  solve the Schr\"{o}dinger equation for a relativistic  particle in a one dimensional-box. Then, we find the energy spectrum and the states of the particle.   \\
\indent
This paper is organized as follows: In section 2, we review the basic elements of the new theory for a one-particle one-dimensional microsystem. In section 3, we find the relativistic equation of motion for a PF system and in section 4, we show that it is invariant under Lorentz transformation. In section 5, using the basic elements of the theory depicted in \cite{shafiee},
we will derive relativistic Schr\"{o}dinger equation for a one-dimensional conservative system, considering both cases of
mass-dependent and mass-independent potentials.  Then in section 6, we will find the state and the energy of a
relativistic particle in a one-dimensional box  to show that the result is physically intuitive and consistent.
At the end, in section 7,  the whole content of our paper is discussed and concluded.
%%%%%%%%%%%%%%%%%%%%%%%%%%%%%%%%%%%%%%%%%%%%%%%%%%%%%%%%%%
%%%%%%%%%%%%%%%%%%%%%%%%%%%%%%%%%%%%%%%%%%%%%%%%%%%%%%%%%%
\section{ Review of Basic Elements}
In this section, we give a brief review of basic elements of the PF theory. More details are available in \cite{shafiee}. For a one-dimensional, one-particle microsystem, three physical entities are introduced:\\
\indent
1. A particle with mass $m$ and position $x(t)$ whose dynamics is given by the Newton's second law:
\begin{equation}
m\frac{d^{2}x(t)}{dt^{2}}=f_{P},
\end{equation}
where $f_{P}$ is the force defined for the particle. For the conservative forces, the particle possesses a conserved energy $E_{P}=V_{P}+K_{P}$, where $K_{P}=\frac{p^{2}_{P}}{2m}$ is the kinetic energy and $p_{P}$ is the linear momentum of the particle.\\
\indent
2. Like the particle aspect of the PF system, there is a field denoted by $X(x(t),t)$  with velocity
 $v_{F}=|\frac{dX}{dt}|=|\dot{X}|$ along the positive direction of $x$, where
%سرعت میدان بررسی شود که دقیقا چه مفهومی دارد مثل سرعت موج است؟
\begin{equation}\label{}
\dot{X}=\left(\frac{\partial X}{\partial x}\right)v_{P}+\left(\frac{\partial X}{\partial t}\right),
\end{equation}
and $v_{P}$ is the velocity of the particle along the same direction. The amplitude of the field has a dimension of length.
Similar to the particle, we assume that the field obeys a Newton-like dynamics too in the same direction,
\begin{equation}\label{a34}
m\frac{d\dot{X}}{dt}=f_{F},
\end{equation}
where $f_{F}$  is the force the field is subjected to. If the particle is subjected to a conservative force $f_{P}$, we shall consider $X=\chi(x(t))$. Then, one can show that
\begin{equation}\label{a341}
v_{F}=|(\frac{d\chi}{dx})|v_{P}=|\chi^{\prime}|v_{P},
\end{equation}
and
\begin{equation}\label{a35}
f_{F}=mv^{2}_{P}\frac{d|\chi^{\prime}|}{dx}+|\chi^{\prime}|f_{P}.
\end{equation}
From a physical point of view, the field $X$ merely enfolds the particle. It experiences its own mechanical-like force introduced as $f_{F}$ in (\ref{a34}), although the presence  of the particle is essential for defining the force of the field. If there is no particle, there will not be any associated field too. The existence of the field depends on the existence of the particle, but the opposite is not true, because $X$ is a function of the particle's position, not vice versa.\\
\indent
For a conservative field subjected to the force $f_{F}$ in (\ref{a35}), one can define the energy $E_{F}=V_{F}+K_{F}$ where $K_{F}=\frac{1}{2}mv^{2}_{F}=K_{P}|\chi^{\prime}|^{2}$. The kinetic energy of the field includes the kinetic energy of the particle. Here, one can't separate the meaning of  $K_{F}$ from $K_{P}$.\\
\indent
In the quantum domain,  the quantities $E_{P}$ and $E_{F}$ are not practically discernible, but the total energy $E=E_{P}+E_{F}$
is an observable property. One can write the total energy as:
\begin{eqnarray}\label{a36}
E&=&V_{P}+(E_{F}+\frac{p^{2}_{P}}{2m}),\nonumber \\
&=&V_{P}+\frac{p^{2}}{2m}
\end{eqnarray}
where $\frac{p^{2}}{2m}=(E_{F}+\frac{p^{2}_{P}}{2m})$, and $V_{P}$ is the particle's potential. \\
\indent
3. Neither the particle, nor the field representation alone is adequate for explaining the physical behavior of a microsystem, comprehensively. What really gives us a thorough understanding of the nature of a quantum system is a holistic depiction of both particle and its associated field which we call  here a PF system. The kinetic energy of a PF system is proportional to
$K_{P}+K_{F}$, but its total energy is the same as $E$ in (\ref{a36}).
Let us define the kinetic energy of a PF as $K_{PF}=\frac{1}{2}m\dot{q}^{2}$, where $q$ denotes the position of the PF and $\dot{q}$
is its velocity. Then, it is legitimate to suppose that $K_{PF}\propto K_{P}+K_{F}$, or
\begin{equation}\label{a361}
\dot{q}^{2}=g_{PF}^{2}(\dot{x}^{2}+|\dot{X}|^{2})
\end{equation}
where $g_{PF}$ is a proportionality factor and $\dot{x}=v_{P}$. For many problems, this factor is equal to one, but the non-oneness of its value in general is crucial in some other problems \cite{shafiee,shafiee2}.\\
\indent
The above relation can be rewritten as a geometric relation in Euclidean space:
\begin{equation}\label{a37}
dq^{2}=g_{PF}^{2}(dx^{2}+|dX|^{2})
\end{equation}
From this relation, one can obtain the trajectories of a PF system:
\begin{equation}\label{a38}
q(x,t)=g_{PF}\int dx\sqrt{\left(1+|\frac{dX(x,t)}{dx}|^{2}\right)}.
\end{equation}
\indent
The relation (\ref{a37}) shows that while we expect the particle to move along the infinitesimal displacement $dx$ in the $x$ direction, the displacement of the whole system is equal to $dq$, not $dx$. The difference here is due to the existence of the associated field which adds a new term, in addition to the direction the particle moves along. Hence, the PF system indeed keeps going through an integrated path determined by the whole action of the particle and its associated field.\\
\indent
Using the relation (\ref{a38}), one can obtain the finite displacement $q$ of a PF system in terms of the particle's location $x(t)$ and time, when the field $X(x,t)$ is known. Then, if the form of dependence of $x$ to $t$ is also known for a given physical problem, it is possible to write $q$ totally in terms of $t$. For stationary states, however, $q=q(x(t))$ and there is no explicit time-dependency. Therefore, one can see that the time variable could be kept concealed in equations of motions, so that the spatial direction $x$ would be sufficient for illustrating the behavior of $q$.\\
\indent
The dynamics of the PF system can also be described according to a Newtonian equation. So, we have
\begin{equation}\label{a1b}
m\frac{d^{2}q}{dt^{2}}=f_{PF},
\end{equation}
where $f_{PF}$ is the force the PF system is subjected to. Using the relations (\ref{a341}) and (\ref{a361}),
one can obtain a relation for  $f_{PF}$  in  stationary states (i.e., the states in which $ X=\chi(x(t))$:
\begin{equation}\label{a39}
f_{PF}=g_{PF}\left[f_{P}(1+\chi^{\prime 2})^{\frac{1}{2}}+\frac{m\dot{x}^{2}\chi^{\prime}\chi^{\prime \prime}}{(1+\chi^{\prime 2})^{\frac{1}{2}}}\right],
\end{equation}
\indent
Regarding our description of a PF system, one may pose the question that what the differences are between this approach and Bohmian account for a micro-system. In other words, what is the advantage of the PF description instead of a Bohmian one? \\
Here are some main points:\\
\indent
1. Contrary to Bohmian Mechanics \cite{bohm, holland}, a PF system is not composed of a particle and a wave. Instead, it is a unified system for which the particle and the wave notions are only abstract constructions without real manifestation. A PF system is neither a particle nor a wave, not also a combination of these two entities. It is a totality of both wave and particle notions, so that one can imagine it as a field that enfolds a particle. However, this is only an imagination. We abstract the notions particle and field to describe the PF system more elaborately. So, it seems that particle and field construct the PF system and when the energy of the field approaches zero, the classical particle appears. Yet, in reality, there are no distinct entities such as the particle and the field. Only when the PF system loses its all holistic nature, it reduces to a known classical particle. Thus, it looks like we have two different energies; one for the particle and the other for the field and the latter causes the quantum behavior of the system. \\
\indent
This important feature of a PF system enables one, e.g., to show why the squared modulus of the wave function behaves like a probability density in spatial coordinates. While, in Bohmian theory Born postulate is accepted a priori. There are other important consequences too which are mentioned in the following items. \\
\indent
2. Considering a PF system allows one to explain the origin of the Schrödinger equation, since here we assume that the underlying dynamics of a supposed field is influenced by an oscillatory force which could be approximated to a harmonic one as the first order [1]. Taking into account anharmonic effects, one can obtain non-linear forms of the Schrödinger equation with additional terms which change the amount of energy of the system at a scale smaller than the hyperfine structure. Such new predictions are forbidden in Bohmian approach in which only quantum predictions are reproduced.\\
\indent
3. The PF theory is not in contradiction with Special Relativity in its origin. Since, contrary to Bohmian Mechanics, there is no need to assume faster-than-signaling between the particles in a many-body system. Yet, the theory is an instance of a contextual local model which have holistic features \cite{shafiee, shafiee2}. This helps us to search for a Lorentz-invariant form of equations in PF theory which is the main purpose of this paper. This fact alone is enough to show the importance of this work.\\
\indent
In addition to above points, there are other differences between the Bohmian and the PF approaches. From a fundamental point of view, these two theories explain bizarre quantum phenomena like the measurement problem, tunneling effect and double-slit experiment in completely different directions. The interested reader can follow the corresponding fashions of explanation in each model in \cite{shafiee, shafiee2, shafiee3} and \cite{bohm, holland}.
%
%For simplicity, hereafter, we take $g_{PF}=1$, except for the cases which need more elaboration.\\
\indent
%The relation (\ref{a39}) which demonstrates the significant role of $K_{P}$ in the definition of kinetic energy of the whole system, also helps us to interprete the meaning of X (and so $\psi$) clearly. To reach a coherent meaning of $\chi$, one should note that whenever $K_{PF}$ is minimized at a given location, one can find the enfolded particle more likely there.
%%%%%%%%%%%%%%%%%%%%%%%%%%%%%%%%%%%%%%P%%%%%%%%%%%%%%%%%%%%%%%%%%%%%%%%%%%%%%%%%%%%%%%%%%
%%%%%%%%%%%%%%%%%%%%%%%%%%%%%%%%%%%%%%%%%%%%%%%%%%%%%%%%%%%%%%%%%%%%%%%%%%%%%%%%%%%%%%%%%
%%%%%%%%%%%%%%%%%%%%%%%%%%%%%%%%%%%%%%%%%%%%%%%%%%%%%%%%%%%%%%%%%%%%%%%%%%%%%%%%%%%%%%%%%
\section{Relativistic Equation of Motion for the PF System  }
%لیبلها رقم سمت چپ a یعنی اولین مقاله
%توضیحات بیشتری در ابتدا داده شود و یک ربط به مفاهیم پایه بیان شود
In this section we are going to find a relativistic form of the relation. (\ref{a38}), to reach a unified concept of spacetime that is invariant under Lorentz transformation.\\
\indent
With the definition of the kinetic  energy of the conservative field, $K_{F}=K_{P}\chi^{\prime 2}$, we define the relativistic
 kinetic energy of the stationary field as
\begin{equation}\label{a1}
K_{rF}=K_{rP}\chi^{\prime 2}
\end{equation}
where $K_{rP}=m_{0}c^{2}(\gamma_{p}-1)$	 is  the relativistic energy of the particle and  $c$  is the velocity of light.
So,  one can define the kinetic energy of the PF system as
\begin{eqnarray}\label{a2}
K_{rPF}&=&K_{rP}+K_{rF}\nonumber \\
&=&m_{0}c^{2}(\gamma_{PF}-1).
\end{eqnarray}
Here
\begin{equation}\label{a3}
\gamma_{PF}=(1-\frac{\dot{q}^{2}}{c^{2}})^{-\frac{1}{2}}
\end{equation}
where $\dot{q}^{2}$ is the velocity of the PF system and $\dot{q}\leq c$. Using the relations  $(\ref {a1})-(\ref {a3})$, we can express an explicit
relation for $\dot{q}$  as the following:
\begin{equation}\label{a4}
\dot{q}=c\left( 1-\frac{1}{[(\gamma_{p}-1)(1+\chi^{\prime 2})+1]^{2}} \right)^{-\frac{1}{2}}.
\end{equation}
Relation (\ref{a4}) shows clearly that for photons, (for which $\gamma_{P}\rightarrow \infty$),  $v_{P}=\dot{q}=c$.\\
\indent
We know that  two infinitesimally separated points $(ct, q)$ and $(ct+cdt, q+dq)$ can be connected by a light signal, according to
the following relation:
\begin{equation}\label{a5}
ds^{2}=c^{2}dt^{2}-dq^{2}.
\end{equation}
This introduces a unified concept of spacetime in the microworld.  In the following, we show that  (\ref{a5}),  defined as the proper
distance, is invariant under Lorentz transformation.% i.e it  has the same form in frames of reference which are moving relative to each other with a constant uniform velocity $\overrightarrow{v}$.
%The transformations between such frames according to the theory of Special Relativity are described by Lorentz transformations.
%
%%%%%%%%%%%%%%%%%%%%%%%%%%%%%%%%%%%%%%%%%%%%%%%%%%%%%%%%%%%%%%%%%%%%%%
%%%%%%%%%%%%%%%%%%%%%%%%%%%%%%%%%%%%%%%%%%%%%%%%%%%%%%%%%%%%%%%%%%%%%%
%%%%%%%%%%%%%%%%%%%%%%%%%%%%%%%%%%%%%%%%%%%%%%%%%%%%%%%%%%%%%%%%%%%%%%
% تعمیم نسبیتی و اینکه چرا ناوردایی لورنتس را حفظ می کند
%عنوان این قسمت ؟؟؟؟؟؟
\section{Lorentz Invariance of the Proper Distance for a PF System}
We can verify whether the relation (\ref{a5}) is invariant under Lorentz transformation in the PF theory,  i.e. , it  has the same form in frames of reference which are moving relative to each other with a constant uniform velocity, $\textbf{v}_{PF}$. Here  $\textbf{v}_{PF}$ shows the relative velocity of two reference frames  corresponding to  the PF system. Let  $Q$  and $Q^{\prime}$  be reference frames for the coordinate systems $(t, q_{x}, q_{y}, q_{z}) $ and $(t^{\prime},q_{x}^{\prime}, q_{y}^{\prime}, q_{z}^{\prime})$, respectively. Without loose of generality, we concern ourselves with the case that the corresponding axes are aligned, with  $q_{x}$ and $q_{x}^{\prime}$  along the line of the relative motion, so that $Q^{\prime}$ has velocity $v_{PF}$ in $q_{x}$  direction in the reference frame of  $Q$. Moreover, we assume that the origins of coordinates and time are chosen so that the origins of  two reference frames coincide at $t=t^{\prime}=0$. Hereafter, we refer to this arrangement as the "standard  configuration" of a pair of reference frames.\\
\indent
 In such a standard configuration, if an event has coordinates $(t, q_{x}, q_{y}, q_{z}) $ in $Q$, then its coordinates in
 $Q^{\prime}$  are given by
%معنی دقیق event
\begin{eqnarray}\label{a6}
q_{x}^{\prime}&=&\frac{q_{x}-v_{PF}t}{\sqrt{1-\frac{v_{PF}^{2}}{c^{2}}}}\\ \nonumber
q_{y}^{\prime}&=&q_{y}\\ \nonumber
q_{z}^{\prime}&=&q_{z}\\ \nonumber
t^{\prime}&=&\frac{t-\frac{v_{PF}}{c^{2}}q_{x}}{\sqrt{1-\frac{v_{PF}^{2}}{c^{2}}}}
\end{eqnarray}
where $v_{PF}=\dot{q}$.
To clarify the approach followed here, let us remember the  entities explained in section 2. Three physical entities have been
defined, a particle with mass $m$,  position $x(t)$ and velocity $v_{P}=\frac{dx}{dt}=\dot{x}$. Associated with the particle, we have
a  field denoted by $\chi(x)$ with the velocity $v_{F}=\frac{d\chi}{dt}=|\chi^{\prime}|v_{P}$ (here,  we are considering the stationary
fields).  The whole PF-system is characterized by the position $q$ and  the  velocity $ v_{PF}=\frac{dq}{dt}=\dot{q}$. \\
%این جمله تصحیح شود
%
\indent
Using the relation (\ref{a4}),  the proper distance defined in (\ref{a5}) is obtained as
\begin{eqnarray} \label{a9}
ds^{2}&=&c^{2}dt^{2}-dq^{2}=c^{2}dt^{ 2}[ 1-\frac{1}{ c^{2}}(\frac{dq}{dt})^{2}]\\ \nonumber
&=&c^{2}dt^{ 2}( \frac{1}{[ (\gamma_{p}-1)(1+\chi^{\prime 2})+1]^{2} }),
\end{eqnarray}
where $\gamma_{p}=\frac{1}{\sqrt{1-\frac{v_{p}^{2}}{c^{2}}}}$ and $\chi^{\prime}=\frac{d\chi}{dx}$\footnote{If $\chi \rightarrow 0$,  the relation (\ref{a9})
will go to  $ds^{2}=c^{2}dt^{2}-dx^{2}=c^{2}dt^{2}\gamma^{-2}_{p}=c^{2}dt^{2}(1-\frac{v^{2}_{p}}{c^{2}})$.}
.
It is easy to  obtain (\ref{a9})  in the primed frame as
\begin{eqnarray}\label{a10}
ds^{ 2}&=&c^{2}\gamma^{2}_{PF}(\frac{v_{PF}}{c^{2}}dq^{\prime}+dt^{\prime})^{2}-\gamma^{2}_{PF}(dq^{\prime}+v_{PF}dt^{\prime})^{2}\nonumber\\
&=&c^{2}\gamma^{2}_{PF}dt^{\prime 2}(1+\frac{v_{PF}}{c^{2}}\frac{dq^{\prime}}{dt^{\prime}})^{2}-\gamma^{2}_{PF}dt^{\prime 2}(v_{PF}+\frac{dq^{\prime}}{dt^{\prime}})^{2}\nonumber\\
&=&c^{2}dt^{\prime 2}[1-\frac{1}{c^{2}}(\frac{dq^{\prime}}{dt^{\prime}})^{2}]\nonumber\\
&=&c^{2}dt^{\prime 2}[ (\gamma_{PF} \frac{1-\frac{v_{p}v_{PF}}{c^{2}}}{\sqrt{1-\frac{v_{p}^{2}}{c^{2}}}}-1)(1+\gamma^{2}_{PF}(\frac{d\chi}{dx^{\prime}}-\frac{v_{PF}}{c^{2}}\frac{d\chi}{dt^{\prime}})^{2})+1]^{-2}
\end{eqnarray}
where we have used the following equations:
\begin{subequations}
\begin{eqnarray}
dt&=&\gamma_{PF}(\frac{v_{PF}}{c^{2}}dq^{\prime}+dt^{\prime}), \\
dq&=&\gamma_{PF}(v_{PF}dt^{\prime}+dq^{\prime}),\\
\chi^{\prime}&=&\frac{d\chi}{dx}=\gamma_{PF}(\frac{d\chi}{dx^{\prime}}-\frac{v_{PF}}{c^{2}}\frac{d\chi}{dt^{\prime}}) \nonumber \\
&=&\gamma_{PF}\frac{d\chi}{dx^{\prime}}(1-\frac{v_{PF}v^{\prime}_{p}}{c^{2}}),
\end{eqnarray}
\end{subequations}
and $\gamma_{PF}$ is defined in (\ref{a3}). The  relation (\ref{a10}) is  the  infinitesimal interval  of two events  in the frame $Q^{\prime}$
\footnote{We see that if the field character $\chi$ vanishes , relation (\ref{a10})  will lead to
 to $ds^{\prime2}=c^{2}dt^{\prime2}-dx^{\prime2}=c^{2}dt^{\prime2}{\gamma^{\prime}_{p}}^{-2}
 =c^{2}dt^{\prime2}(1-\frac{v^{\prime2}_{p}}{c^{2}})$  },
It  seems that the form of the relations   (\ref{a10}) and  (\ref{a9}) are different in the two frames $Q$ and $Q^{\prime}$ which
are moving relative to each other with the constant velocity $\textbf{v}_{PF}$.\\
 % For invariance under Lorentz transformation two equation in the $s$ and $s^{\prime}$ must be the same.\\
\indent
Now,  we try to find  the constant $\gamma_{PF}$ so that the  relations   (\ref{a10}) and  (\ref{a9}) have
 the same form in the two reference frames. It is worth to  note that   $\gamma_{PF}$  is a function  of
 $v_{P}$, $v^{\prime}_{P}$, $\frac{d\chi}{dt}$, $\frac{d\chi}{dt^{\prime}}$, provided  that   when $\chi \rightarrow 0$,
 we get the following relations  between the gamma factors:
% ,$v_{P}$,$v^{\prime}_{P}$ and$v$ , as
%
\begin{equation}\label{a8}
\frac{1}{\sqrt{1-\frac{v_{P}^{2}}{c^{2}}}}=\frac{1+\frac{v^{\prime}_{P}v}{c^{2}}}{\sqrt{1-\frac{v^{2}}{c^{2}}}\sqrt{1-\frac{v_{P}^{\prime2}}{c^{2}}}}.
\end{equation}
where $v_{P}$ and $v^{\prime}_{P}$  are the velocities of the particle in the frames $Q$ and $Q^{\prime}$, respectively.\\
%منظور از ذرات کلاسیکی همان تعریف معمول ذره بدون درنظر گرفتن میدان است؟؟؟؟؟؟؟؟؟؟؟؟
\indent
It will be useful to write the relation (\ref{a9}) as
\begin{eqnarray} \label{a11}
ds^{ 2}&=& \frac{c^{2}dt^{ 2}}{[ \gamma_{p}-\chi^{\prime 2}+\gamma_{p}\chi^{\prime 2}]^{2} } \\ \nonumber
&=& \frac{c^{2}dt^{ 2}}{ \gamma_{p}^{2}[1+\chi^{\prime 2}\frac{(\gamma_{p}-1)}{\gamma_{p}}]^{2} }
\end{eqnarray}
and the relation (\ref{a10})  as
\begin{eqnarray}\nonumber
ds^{ 2}&=&c^{2} dt^{ \prime 2}( \frac{1}{[ (\gamma_{PF} a-1)(1+\gamma^{2}_{PF}b^{2})+1]^{2} }) \nonumber\\
&=&c^{2} dt^{ \prime 2} \frac{1}{(\gamma_{PF} a-\gamma^{2}_{PF}b^{2}+\gamma^{3}_{PF}b^{2}a)^{2} } \nonumber\\
&=&\frac{c^{2} dt^{ \prime 2}}{(\gamma_{PF} a-\gamma^{2}_{PF}b^{2}+\gamma^{3}_{PF}b^{2}a)^{2}}\nonumber \\
&=&\frac{c^{2} dt^{ \prime 2}}{\gamma^{2}_{PF}a^{2}[1+b^{2}\gamma^{2}_{PF}(\frac{\gamma_{PF} a-1}{\gamma_{PF} a})]^{2}} \label {14a}
\end{eqnarray}
where $a$ and $b$ are respectively  defined  as
\begin{subequations}
\begin{eqnarray}
a&=&\frac{1-\frac{v_{p}v_{PF}}{c^{2}}}{\sqrt{1-\frac{v_{p}^{ 2}}{c^{2}}}}=\gamma_{p}(1-\frac{v_{p}v_{PF}}{c^{2}}) \label{15a}\\
b&=&\frac{d\chi}{dx^{\prime}}-\frac{v_{PF}}{c^{2}}\frac{d\chi}{dt^{\prime}} \nonumber\\
&=&\frac{d\chi}{dx^{\prime}}(1-\frac{v_{PF}v^{\prime}_{p}}{c^{2}}).\label{a15}
\end{eqnarray}
\end{subequations}
The value of  $\chi$ is small enough  to write  the relations (\ref{a9}) and (\ref{a10}) as \cite{shafiee},
%تقریب دو جمله
\begin{eqnarray}\label {a12}
ds^{ 2}&=&c^{2} dt^{ 2}\gamma_{p}^{-2}[1-2\chi^{\prime 2}\frac{(\gamma_{p}-1)}{\gamma_{p}}+
...]
\end{eqnarray}
%
%in the limit of speed of light, $\gamma_{p}>>1$ and we can write relation (\ref{a12}) in a simpler form
%
%\begin{eqnarray}\label {a13}
%ds^{ 2}&=&c^{2} dt^{ 2}\gamma_{p}^{-2}[1-2\chi^{\prime 2}+
%...]
%\end{eqnarray}
and
\begin{eqnarray}\label {a16}
ds^{ 2}&=&c^{2} dt^{\prime 2}(a\gamma_{PF})^{-2}[1-2b^{2}\gamma^{2}_{PF}\frac{(\gamma_{PF} a-1)}{\gamma_{PF} a}+
%3\frac{b^{4}}{a^{2}}(\gamma^{2}a-\gamma)^{2}+
...]\nonumber\\
&=&c^{2} dt^{\prime 2}(a\gamma_{PF})^{-2}[1-2b^{2}\gamma^{2}_{PF}+...].
\end{eqnarray}
For large values of $v_{P}$, $\gamma_{p}>>1$ and we can write the relation (\ref{a12}) in a simpler form
\begin{eqnarray}\label {a13}
ds^{ 2}&\simeq&c^{2} dt^{ 2}\gamma_{p}^{-2}[1-2\chi^{\prime 2}+
...].
\end{eqnarray}
So, we can find  $\gamma_{PF}$,  such that the  relations  (\ref{a13}) and (\ref{a16})  have the same form under
Lorentz transformation, i.e.,
\begin{eqnarray}
{\gamma_{p}^{\prime}}^{-2}[1-2(\frac{d\chi}{dx^{\prime}})^{ 2}]&=&(\gamma_{PF} a)^{-2}[1-2b^{2}\gamma^{2}_{PF}] \nonumber \\
\gamma^{2}_{PF}a^{2}[1-2(\frac{d\chi}{dx})^{ 2}]&=&\gamma_{p}^{\prime2}[1-2b^{2}\gamma^{2}_{PF}].
\end{eqnarray}
Substituting $a$ and $b$ from the  relations   (\ref{15a}) and  (\ref{a15}), one gets
\begin{equation}\label{a118}
\gamma^{2}_{PF}\gamma_{p}^{2}(1-\frac{v_{p}v_{PF}}{c^{2}})^{2}[1-2(\frac{d\chi}{dx^{\prime}})^{ 2}]=\gamma_{p}^{\prime2}[1-2\gamma^{2}_{PF}(\frac{d\chi}{dx^{\prime}})^{2}
(1-\frac{v^{\prime}_{p}v_{PF}}{c^{2}})^{2}].
\end{equation}
So, we have:
\begin{equation}\label{a18}
\gamma^{2}_{PF}[(1-2(\frac{d\chi}{dx^{\prime}})^{ 2})(1-\frac{v_{p}v_{PF}}{c^{2}})^{2}\gamma_{p}^{2}+2(\frac{d\chi}{dx^{\prime}})^{2}
(1-\frac{v^{\prime}_{p}v_{PF}}{c^{2}})^{2}]
=\gamma_{p}^{\prime2}.
\end{equation}
Then
\begin{equation}
\gamma^{2}_{PF}=\frac{\gamma_{p}^{\prime2}}{[(1-2(\frac{d\chi}{dx^{\prime}})^{ 2})(1-\frac{v_{p}v_{PF}}{c^{2}})^{2}\gamma_{p}^{2}+2(\frac{d\chi}{dx^{\prime}})^{2}
(1-\frac{v^{\prime}_{p}v_{PF}}{c^{2}})^{2}]},
\end{equation}
or
\begin{equation}\label{a19}
\gamma_{PF}=\frac{\gamma^{\prime}_{p}}{[(1-2(\frac{d\chi}{dx^{\prime}})^{ 2})(1-\frac{v_{p}v_{PF}}{c^{2}})^{2}\gamma_{p}^{2}+2(\frac{d\chi}{dx^{\prime}})^{2}
(1-\frac{v^{\prime}_{p}v_{PF}}{c^{2}})^{2}]^{\frac{1}{2}}}.
\end{equation}
%if $\chi \rightarrow 0$, the plus sign gives the correct answer.
It is convenient to write the relation  (\ref{a19}) as
\begin{eqnarray}\label{a20}
\gamma_{PF}&=&\frac{\gamma^{\prime}_{p}}{[(1-\frac{2}{v^{\prime2}_{p}}(\frac{d\chi}{dt^{\prime}})^{ 2})(1-\frac{v_{p}v_{PF}}{c^{2}})^{2}{\gamma_{p}}^{2}+\frac{2}{v^{\prime2}_{p}}(\frac{d\chi}{dt^{\prime}})^{2}(1-\frac{v^{\prime}_{p}v_{PF}}{c^{2}})^{2}]^{\frac{1}{2}}}\nonumber\\
&=&\frac{\gamma^{\prime}_{p}}{[(1-2\frac{{v^{\prime}_{F}}^{2}}{v^{\prime2}_{p}})(1-\frac{v_{p}v_{PF}}{c^{2}})^{2}{\gamma_{p}}^{2}+2\frac{{v^{\prime}_{F}}^{2}}{v^{\prime2}_{p}}(1-\frac{v^{\prime}_{p}v_{PF}}{c^{2}})^{2}]^{\frac{1}{2}}},
\end{eqnarray}
where $v_{F}=\frac{d\chi}{dt}=v_{p}|\frac{d\chi}{dx}|$  and $v^{\prime}_{F}=\frac{d\chi}{dt^{\prime}}=v_{p}^{\prime}|\frac{d\chi}{dx^{\prime}}|$  are the velocities of the field in two reference frames $Q$ and $Q^{\prime}$,  respectively.
%we see that this equation gives these conditions\\
%1.It has real value \\
In   the relation (\ref{a20}), when $\chi \rightarrow 0$,  we obtain $\gamma^{2}a^{2}=\gamma_{p}^{\prime2}$, which as we expect, is the exact
relationship between the gamma factors in  the relation  (\ref{a8}). In fact,  in this limit,  the PF system converts to the classical particle. \\
\indent
 From the relation  (\ref{a36}), it is obvious that for large values of $v_{P}$,  the  energy  of the particle has the most contribution to the energy of  the PF system. One  can also  see this fact  from  Eq.  (\ref{a20}).  For large values of $v_{P}$ ,
we find  the relation  (\ref{a8}) i.e.,  PF system transforms to the classical particle.
%The relation  (\ref{a20}) shows that  when the velocity  of the particle increases, the amplitude of the field decrease and again  the equation  (\ref{a8}) is obtained.
So, it seems that if  $\gamma_{PF}$ satisfies the relation (\ref{a20}), infinitesimal distance of two separated points, $ds^{2}$, defined in the PF theory  is invariant under Lorentz transformations.
%%%%%%%%%%%%%%%%%%%%%%%%%%%%%%%%%%%%%%%%%%%%%%%%%%%%%%%%%
%%%%%%%%%%%%%%%%%%%%%%%%%%%%%%%%%%%%%%%%%%%%%%%%%%%%%%%%%
%%%%%%%%%%%%%%%%%%%%%%%%%%%%%%%%%%%%%%%%%%%%%%%%%%%%%%%%%
%%%%%%%%%%%%%%%%%%%%%%%%%%%%%%%%%%%%%%%%%%%%%%%%%%%%%%%%%
%%%%%%%%%%%%%%%%%%%%%%%%%%%%%%%%%%%%%%%%%%%%%%%%%%%%%%%%%
\section{Relativistic Generalization of  Time Independent Schr\"{o}dinger Equation}
% تعمیم نسبیتی معادله شرودینگر صفحه 21 تا معادله 68ت هست. در اینجا صحبتی از اسپین نشده است.
Now, we are going to derive a relativistic  Schr\"{o}dinger equation in a general form. The physical structure of the PF formalism which has constitutional similarities to classical equations of motion permits us to derive a well-defined relativistic Schr\"{o}dinger equation for stationary fields, regardless of spin variable.
The dynamics of a stationary real field in one dimension (denoted by $\chi=\chi(x(t))$ in the relativistic regime can be represented as
\begin{equation}
\frac {d(m_{p}\dot{\chi})}{dt}=f_{rF},
\end{equation}
where $f_{rF}$ is the force defined for the field under the relativistic conditions  and  $m_{p}$ is the relativistic mass of
the particle:
\begin{equation}\label{a23}
m_{p}=\gamma_{p}m_{0};\hspace{1cm}\gamma_{p}=\left(1-\frac{v^{2}_{p}}{c^{2}}\right)^{-\frac{1}{2}},
\end{equation}
where $m_{0}$  is the rest mass, as before. The stationary field $\chi(x(t))$ does not explicitly depend on time. So, one can find out that
\begin{equation} \label{a24}
f_{rF}=f_{rP}\chi^{\prime}+\gamma_{P}m_{0}v^{2}_{P}\chi^{\prime\prime}
\end{equation}
where $\chi^{\prime}=\frac{d\chi}{dx}$ and $f_{rP}=m_{0}\dot{v}_{P}\gamma_{P}^{3}$ is the force exerted on the particle. For stationary real fields, there exists an oscillating-like term in the force expression (denoted by the second term in (\ref{a24}), when $\gamma_{P} \rightarrow 1$) from which the non relativistic  time-independent Schr\"{o}dinger equation can be resulted
\cite{shafiee}.
Here, we suppose that the same situation holds true under the relativistic conditions. That is, for stationary real fields, we postulate the following equality as a general rule:
\begin{equation}\label{a25}
-m_{P}\bar{w}^{2}\chi= \gamma_{P}m_{0}v^{2}_{P}\chi^{\prime\prime},
\end{equation}
where $m_{P}$ was defined in relation (\ref{a23}) and $\bar{w}^{2}=k^{2}v^{2}_{P}$. Here again, we define $k=\frac{p}{\hbar}$,
where p is the relativistic de Broglie momentum. Fromthe relation  (\ref{a25}), it is immediately concluded that
\begin{equation}\label{a26}
-\hbar^{2}\chi^{\prime\prime}=p^{2}\chi,
\end{equation}
which has the same form as the non-relativistic Schr\"{o}dinger equation. To find an appropriate relation for $p^{2}$ in relation (\ref{a26}), however,  one should first note that depending on whether the potential energy of the particle is a function of mass or not\footnote{Sometimes expressed inversely, that is whether the mass is potential-dependent or not \cite{Harvey}},
the total energy of the PF system can be respectively written as
\begin{eqnarray}\label{a27}
E&=&\gamma_{P}(V_{nrP}+m_{0}c^{2})+E_{rF}\nonumber \\
&=&\gamma(V_{nrP}+m_{0}c^{2})
\end{eqnarray}
or
\begin{eqnarray}\label{a28}
E&=&V_{nrP}+\gamma_{P}m_{0}c^{2}+E_{rF}\nonumber \\
&=&V_{nrP}+\gamma m_{0}c^{2}.
\end{eqnarray}
where $V_{nrP}$ is the non relativistic potential energy of the particle and $E_{rF}$ is the relativistic energy of the field. Also, using the relation $p=\gamma m_{0}v$, $\gamma$ can be defined as
\begin{equation}\label{a29}
\gamma=\left( 1-\frac{v^{2}}{c^{2}}\right)^{-\frac{1}{2}}=\sqrt{1+\frac{p^{2}}{m_{0}^{2}c^{2}}}
\end{equation}
At the nonrelativistic limit, we take $\gamma V_{nrP}\rightarrow V_{nrP}$,
$\gamma m_{0}c^{2}\rightarrow \frac{p^{2}}{2m_{0}}+m_{0}c^{2}$ and $E_{rF}\rightarrow E_{nrF}$,
so that the total energy in relations  (\ref{a27}) or (\ref{a28})  can be expressed as
%
%\begin{equation}
$$E= V_{nrP}+ \frac{p^{2}}{2m_{0}}+m_{0}c^{2}= E_{nr}+m_{0}c^{2},$$
%\end{equation}
%
where $p^{2}=p^{2}_{P}+2m_{0}E_{nrF}$. From  relations (\ref{a27}) and (\ref{a29}),  we can deduce that
\begin{equation}\label{a30}
p^{2}=\left(1+\frac{V_{nrP}}{p^{2}}\right)^{-2}\left[\frac{E^{2}}{c^{2}}-m_{0}^{2}c^{2}\left(1+\frac{V_{nrP}}{p^{2}}\right)^{2}\right]
\end{equation}
In a similar manner, from  (\ref{a28}) and (\ref{a29}) ,  one obtains
\begin{equation}\label{a31}
p^{2}=\frac{1}{c^{2}}(E-V_{nrP})^{2}-m_{0}^{2}c^{2}
\end{equation}
Inserting the relation (\ref{a30}) in  (\ref{a26}), we derive  relativistic Schr\"{o}dinger equations for the cases that the
potential energy of the  particle includes the relativistic mass:
\begin{equation}\label{a32}
-\frac{\hbar^{2}}{2m_{0}}\chi^{\prime\prime}+\frac{1}{2}m_{0}c^{2}\chi=
\frac{E^{2}}{2m_{0}c^{2}}\left(1+\frac{V_{nrP}}{m_{0}c^{2}}\right)^{-2}\chi,
\end{equation}
and using (\ref{a31}), we  get the relativistic Schr\"{o}dinger equation, when the potential energy of
the particle is independent of mass:
\begin{equation}\label{a33}
-\frac{\hbar^{2}}{2m_{0}}\chi^{\prime\prime}+\frac{1}{2}m_{0}c^{2}\chi=\frac{1}{2m_{0}c^{2}}(E-V_{nrP})^{2}\chi.
\end{equation}
So,  one can solve relations (\ref{a32}) or (\ref{a33}) for different problems.
In the following,  we consider the problem of a particle in a one-dimensional box to find its eigenstates and its energy spectrum.
We solve such equations  for two  other problems, one-dimensional harmonic oscillator and the relativistic Hydrogen
in a separate article.
%%%%%%%%%%%%%%%%%%%%%%%%%%%%%%%%%%%%%%%%%%%%%%%%%%%%%%%%%%
%%%%%%%%%%%%%%%%%%%%%%%%%%%%%%%%%%%%%%%%%%%%%%%%%%%%%%%%%%
%%%%%%%%%%%%%%%%%%%%%%%%%%%%%%%%%%%%%%%%%%%%%%%%%%%%%%%%%%
%%%%%%%%%%%%%%%%%%%%%%%%%%%%%%%%%%%%%%%%%%%%%%%%%%%%%%%%%%
\section{Relativistic Micro-Particle in One-Dimensional Box}
The simplest system which could be analyzed is a PF system in a one-dimensional box. We consider a particle for which the
nonrelativistic potential energy is defined as
\begin{eqnarray}
V_{nrP}&=&0 \hspace{1cm} 0\leq x \leq a\\ \nonumber
V_{nrP}&=&\infty \hspace{1cm} elsewhere.
\end{eqnarray}
So the relation (\ref{a32}) can be written as
\begin{equation}\label{a331c}
\chi^{\prime\prime}(x)=-k^{2}\chi(x),
\end{equation}
where
\begin{equation}\label{a331}
k^{2}=\frac{1}{\hbar^{2}c^{2}}(E^{2}-m_{0}^{2}c^{4}).
\end{equation}
At the boundaries,  we have,
 $$\chi(0)=\chi(a)=0.$$
We  solve the relativistic Schr\"{o}dinger equation (\ref{a33})
 to obtain the energy spectrum and the stationary eigenfields,  respectively,  as,
\begin{equation}\label{a331b}
(\frac{E_{n}}{c})^{2}=\frac{n^{2}h^{2}}{4a^{2}}+m_{0}^{2}c^{2}.
\end{equation}
 and
\begin{equation}\label{a332}
\chi_{n}(x)=A_{rn}\sin(\frac{n\pi x}{a}); n=1, 2, 3, ...
\end{equation}
where  $A_{rn}$  is the amplitude of the field which can be found, if the relativistic energy of the stationary field $E_{rF}$ is known. The trajectories of the PF system can be obtained by integrating (\ref{a4}) over t,  but since $\chi_{n}^{\prime}(x)$ is a function of $x(t)$, the solution is complicated.\\
\indent
For a photonic PF system in one-dimensional box, $E_{n}=\frac{nhc}{2a}$. The energies of the  particle
$E_{rP}=\gamma_{P}m_{0}c^{2}$ and its associated field $E_{rF}$
are finite but unknown. The de Broglie momentum of the photonic PF system is also sharp around the values $\pm\frac{nh}{2a}$.\\
\indent
For a free photonic PF system, we have the same relation as (\ref{a331}), but with  $E=pc$ and $p=\frac{h}{\lambda}$
where $\lambda$ is the wavelength. The energy of a free photon is not quantized, but it can be still described as
a PF system comprised of a particle and its allied field which the latter behaves like a plane wave and the whole
system propagates with velocity c. The trajectories of the free photonic PF system are straight lines (in terms of t),
because $\dot{q}=c$ in (\ref{a4}), regardless of the form of $\chi$.
%%%%%%%%%%%%%%%%%%%%%%%%%%%%%%%%%%%%%%%%%%%%%%%%%%%%%%%%%%%%
%%%%%%%%%%%%%%%%%%%%%%%%%%%%%%%%%%%%%%%%%%%%%%%%%%%%
%%%%%%%%%%%%%%%%%%%%%%%%%%%%%%%%%%%%%%%%%%%%%%%
%%%%%%%%%%%%%%%%%%%%%%%%%%%%%%%%%%%%
\section{Discussion}
The new formalism of microphenomena introduced in
\cite{shafiee}  can be used to generalize Quantum Mecchanics to include relativistic equations, regardless of the spin notion.\\
\indent
 Using the relativistic kinetic energy of the field and the particle denoted by a PF system, we used a more general,
 equation of motion of  a PF system relation (\ref{a4})
,  to show that an infinitesimally separated point according to  (\ref{a5}) is invariant under Lorentz transformation. Considering a standard  configuration of a pair reference frames, we found the proper distance in both of them. Since in the PF theory a field accompanies the particle, the relative velocity
 of two reference frames is a function of the velocity of  the particle and its field.  So, if $\gamma_{P}$ satisfies the
 the relation  (\ref{a20}), the proper distance is invariant under Lorentz transformation.\\
\indent
 Moreover, considering the classical
 potential energy, we derived the relativistic Schr\"{o}dinger equation for the case that the potential energy of the particle
 includes the relativistic mass and the case that it is independent of it. We solved the relativistic Schr\"{o}dinger equation
 to find the energy spectrum and the stationary eigenfields of a relativistic PF system for the one-dimensional box.\\
\indent It is worth mentioning here that we have not  yet defined the spin which is  one of the conceptual puzzles in Quantum Mechanics. Although there is a consensus about elementary particles having some quantum mechanical property  called spin, the understanding of the physical nature of the spin is still incomplete \cite{morrison}. Historically, the concept of spin was introduced in order to explain some experimental findings such as the emission spectra of alkali metals and Stern-Gerlach experiments. Though the spin is regarded as a fundamental property of the electron, a universally accepted spin operator for the Dirac theory is still missing  \cite{Bauke}.
So, we note that  the relativistic Schr\"{o}dinger equations, i.e., the relations (\ref{a32})  and (\ref{a33}), found  in the relativistic generalization of the PF theory are completely different with Dirac equation, derived for half spin particles. Also it is not the same as  Klein-Gordon equation for spin zero particles. and doesn't have the problem of negative energies. \\
\indent last but not least,  it is important mentioning that the causal basis of the PF theory along with its capability to be reformulated on a geometric ground makes it one think over the general relativistic  development of this new theory in a rigorous way.


\begin{thebibliography}{widest-label}
\bibitem{shafiee}A. Shafiee, %"On a new formulation of microphenomena: Basic principles, stationary fields and beyond",
Pramana journal of physics, {\bf 76}, No. 6, 843-873, (2011).
\bibitem{shafiee2} A. Shafiee, A. Massoudi and M. Bahrami, %"On A New Formulation of Micro-phenomena: The Double-slit Experiment",
quant-ph/arXiv:0810.1034.
\bibitem{shafiee3} A. Shafiee, quant-ph/arXiv:0810.1033.
\bibitem{einstein}L. E. Ballentine, Am. J. Phys. {\bf 40}, (1972).
\bibitem{epr}A. Einstein, B. Podolsky and N. Rozen, Phys. Rev. {\bf 47}, 777 (1935)
%\bibitem{Harvey}A. L. Harvey, %"Relativistic Harmonic Oscillator",
Physical Review D, {\bf 6} 1474-1476, (1972).
\bibitem{bohm}D. Bohm and B. J. Hiley, The Undivided Universe (Routledge, London, 1993).
\bibitem{holland}P. Holland, The Quantum Theory of Motion (Cambridge University Press, New York, 1993).
\bibitem{morrison}M. Morrison, Studies In History and Phylosophy of Science Part B: Studies In History and Philosophy of Modern Phyics {\bf 38}, 529, (2007).
\bibitem{Bauke}H. Bauke, S. Ahrens, C. H. Keitel and R. Grobe,
%"What is relativistic spin operator"
New Journal of Physics, {\bf 16}, 043012, (2014).
\end{thebibliography}
\end{document}